# Sustaining educational and public outreach programs in astronomy

Will Clarkson[1], Don Bord[1], Carrie Swift[1], Eric J. Rasmussen[1], Dave Matzke[1],
Steve Murrell[2], Mike LoPresto[2],
Timothy Campbell[3], Robert Clubb[3], Dennis Salliotte[3]

[1]University of Michigan-Dearborn, 4901 Evergreen Road, Dearborn, MI 48128; wiclarks@umich.edu
[2]Henry Ford College, 5101 Evergreen Rd, Dearborn, MI 48128
[3]The Ford Amateur Astronomy Club: http://www.fordastronomyclub.com/index.html

*Abstract:* We advocate meaningful support of sustained education-outreach partnerships between regional metropolitan undergraduate institutions and astronomical clubs and societies. We present our experience as an example, in which we have grown a partnership between the University of Michigan-Dearborn (hereafter UM-D, a 4-year primarily undergraduate institution or PUI), Henry Ford College (hereafter HFC, a 2-year undergraduate college), and maintained a strong collaboration with the Ford Amateur Astronomy Club (FAAC), which is highly active in the Detroit Metropolitan Area. By allowing each organization to play to its strengths, we have developed a continuum of education-outreach efforts at all levels, with connecting tissue between the previously disparate efforts. To-date, faculty and staff effort on these initiatives has been nearly entirely voluntary and somewhat ad-hoc. Here we suggest an initiative to sustain the continuum of education-outreach for the long-term. There are two levels to the suggested initiative. Firstly, partner institutions should dedicate at least half an FTE of faculty or staff effort specifically to education and outreach development. Secondly, professional societies like the AAS now have a great opportunity to support the education-outreach continuum at a national level, by facilitating communication between institutions and clubs that are considering a long-term partnership, by acting as a central resource for such partnerships, and possibly by convening or sponsoring events such as professional meetings among the metropolitan educational community.

*Motivation:* Regional institutions[1] are important economic drivers of the communities in which they reside. With a large regional intake, and (usually) a high proportion of graduates remaining in the geographical area, regional higher-educational institutions are in many areas the face of higher education in the US. Astronomy outreach in particular holds a special place in deepening the community benefits of higher education. For example, at the K-12 level, education/outreach efforts help to change school student preconceptions of the scientist as the "other," and aiding the realization that college is a possibility. Interaction with higher-education students (and faculty) from the same community as these school students provides a success example in science by people like themselves, thus metropolitan regional institutions are particularly effective at this kind of community involvement. Secondly, education/outreach efforts help citizens to exercise their sense of ownership of the public higher education institutions that serve their community, thus maintaining the community-academia link. Importantly, this includes community members seeing the institution as a possible educational destination, which might not have previously been the case. At an individual level, should a participant pursue astronomy as a course of study, they will learn skills that will directly benefit them in the job market (such as data analysis skills or the ability to plan, conduct, and communicate experimental investigation).

A continuum of involvement thus exists, from occasional participation in (or awareness of) public outreach events, through to an individual setting themselves up for a career in research. Many metropolitan areas offer all the pieces along this path, and in nearly all cases the connecting tissue is maintained by volunteer personnel, contributing effort well beyond that for which they are being compensated. We propose a specific initiative to ensure the continuum of effort is formalized and sustained.

*Background:* As an example, we present three institutions within the Detroit Metropolitan Area which have operated together under varying degrees of partnership: a 4-year PUI (UM-D), a 2-year college (HFC), and an amatuer astronomy club (FAAC). Recognizing the degree to which the two institutions complement each other, UM-D and HFC entered into a formalized partnership [1], including the use by students of each others' Astronomical facilities, and the development of educational materials using them. Here follow some relevant characteristics of the three organizations:

- UM-D offers an Astronomy minor as well as lower-level introductory courses, and a small program (~8-10 students) of undergraduate Astronomy research;
- HFC offers highly innovative educational programs, including the development of an Astrobiology diagnostic test analogous to CAER's ACT used in large intro-astronomy courses [2,3];
- Many HFC students will transfer to UM-D to complete their studies
- UM-D operates a campus observatory, with an 0.4m telescope and four smaller telescopes;
- HFC operates a planetarium with 38 seats, which is used in undergraduate education, K-12 outreach, and regularly-scheduled free public programs. It is one of only two planetaria in the area that does not charge for shows. For example, FAAC typically holds 6-7 public planetarium shows *per month* with the HFC planetarium.
- The Ford Amatuer Astronomy Club is prominent in the community with approximately 120 members and offering monthly lecture events, monthly public outreach observing events, and K-12 outreach events each year.
- Both HFC and UM-D have formalized commitment to the Metropolitan Area at high administrative levels see for example the UM-D Mission statement[2] and the HFC Mission statement[2].
- The goal of FAAC is to "encourage the study of Astronomy, math, the sciences, and related subjects for the benefit of its members and the general public" (e.g. FAAC homepage[2]).

___
[1] Here we define "regional institutions" as primarily undergraduate higher education institutions, whether 4-year or 2-year, for whom a majority of incoming freshmen are within driving distance of the institution.

Working together, our three organizations have had substantial impact both within and outside academia. UM-D, HFC and FAAC members together run an informal program of visits by K-12 students from the area, which typically impact large numbers of schoolchildren per year (of order a thousand per year at the planetarium, of order a few hundred per year at the UM-D Observatory). This allows each organization to play to its strengths: for example, most fifth-grade students in the Dearborn area visit the HFC planetarium, UM-D typically hosts about five hundred K-12 students per year in groups of about two dozen, and finally the continuous efforts of FAAC which are conducted year round. One prominent example is the Astronomy at the Beach event, which FAAC co-organizes through its membership in the GLAAC (the Great Lakes Association of Astronomy Clubs), which typically has about 4500 visitors.

*The challenge:* Faculty, staff, and club members traditionally provide these services on a voluntary basis. The effort is substantial [4], including (but not limited to): creation and maintenance of a web presence; administrative effort regarding publicity for events and real-time response to changing weather conditions; correspondence with a large number of individuals and organizations (e.g. local schools); development and maintenance of the facilities used (e.g. the observatory). This leads to at least three important obstacles to maintaining *sustained* involvement:

1. *Time & Personnel:* Sheer time pressure makes maintenance of a program involvement challenging when the effort is purely voluntary. For example, at UM-D the typical semester load is 3 courses per term for full-time faculty, 4 per term for lecturers, leaving little time for extracurricular activities.
2. *Innovation, not just maintenance:* Finding the time to properly develop innovations can thus be challenging. A 0.25 FTE involvement (for example to run and maintain a planetarium) tends in practice not to allow much time to develop innovations.
3. *Mission centricity:* Maintaining continuity can be difficult if the individual maintaining an institution's community education efforts is unavailable (e.g. on medical leave or long term research travel), the program could be compromised. Recognition of education/outreach efforts as a core part of the institutional mission, would also help insulate community involvement from short-term administrative fluctuation at the institutional level.

*Meeting the challenge:* We propose a program to institutionally protect community outreach programs by supporting the individuals who put so much time and effort into it. Specific initiatives:

---
[2] Links: http://umdearborn.edu/about/mission-vision ; https://www.hfcc.edu/about-us/mission ; http://www.fordastronomyclub.com/index.html

*Educational institution:*

1. At least 0.5 FTE support per institution should be committed to an individual whose formal tasks would be to maintain the resource used (e.g. observatory or planetarium), to set up and co-ordinate E/PO events, and to maintain relationships with the community;

2. Outreach efforts should be explicitly recognized as a service category in Promotion and Tenure considerations. For example, creating and conducting some number of community outreach events per semester could be counted equivalently to membership in a single campuswide committee;

3. A long-term community engagement program should be supported, with the form chosen to best-match the partner institution. Example forms might be:
    a. A guest speaker program, where (for example) astronomical society members would host public viewing at the campus observatory, or present a planetarium show;
    b. Establishment of a seminar course for undergraduates, focusing on public understanding of science; in which part of the course requirements is participating in public or K-12 outreach activities;
    c. Internships or co-ops (as a practicum for education students), in which students develop and/or present a K-12 outreach event, or develop new materials for planetarium events;
    d. An events program at the intersection of the amateur Astronomy community and the instructional community. In many cases these may already be in-place (for example UM-D and HFC's involvement with Astronomy at the Beach).

*Professional Society such as the American Astronomical Society (AAS):*

1. We believe that similar efforts and partnerships may be underway at other institutions, although as yet we have not been able to dedicate time to verify this. Professional societies such as the AAS should use their capabilities to gather such data on a national scale, to learn what education/outreach/community partnerships already exist;

2. A substantial amount of effort duplication is likely, since many institutions have some level of community outreach as a component of their core mission. A society like the AAS can provide a repository of knowledge and initiatives in order for the various collaborations to learn from each other's experiences;

3. Convening or sponsoring professional meetings with emphasis on the education/outreach interface.

**Summary:** Using the UM-D + HFC + FAAC collaboration as an example, we argue that co-operative education-outreach efforts between 4-year universities, 2-year colleges, and amateur astronomical clubs, provides important continuity in K-12 educational enhancement and community ownership of higher education and should be meaningfully supported. We identify initiatives that can be taken by partner institutions and professional societies like the AAS, to allow the education-outreach mission of the Astronomical community to be sustained in the long term.